# Efficient and Reversible $CO_2$ Capture by Lithium-functionalized Germanene Monolayer

**S. M. Aghaei** [*], **M. M. Monshi, I. Torres, and I. Calizo**[**]

Department of Electrical and Computer Engineering, Florida International University, Miami, Florida 33172, USA

First-principles density functional theory (DFT) is employed to investigate the interactions of $CO_2$ gas molecules with pristine and lithium-functionalized germanene. It is discovered that although a single $CO_2$ molecule is weakly physisorbed on pristine germanene, a significant improvement on its adsorption energy is found by utilizing Li-functionalized germanene as the adsorbent. However, the moderate adsorption energy at high $CO_2$ coverage predicts an easy release step. More excitingly, the structure of Li-functionalized germanene can be fully recovered after removal of $CO_2$ gas molecules. Our results suggest that Li-functionalized germanene show promise for $CO_2$ sensing and capture with a storage capacity of 12.57 mol/kg.

**1 Introduction** The development of technology for $CO_2$, the main man-made greenhouse gas, capture and storage has garnered huge interest [1, 2]. The current industrial technology which is based on chilled ammonia suffers from toxicity, solvent loss, low energy efficiency, and corrosion issues [3, 4]. To overcome these problems, various materials such as metal-organic frameworks [5], boron nitride nanotubes [6], and carbon nanotubes [7] have been employed as effective sorbents for $CO_2$ capture. Nevertheless, their large adsorption energies lead to a difficult regeneration step [6]. Ergo, the selection of adsorbents with both high selectivity and an easy release step is a vital task.

Since its discovery in 2004 [8], graphene has enticed a great deal of interest due to its outstanding properties. The potential application of graphene for gas sensing has been widely studied [9-12]. Although the physisorption of $CO_2$ on pristine graphene limits its potential for single molecule detection [9], its sensing capability can be improved by modifying graphene [10-12]. It has been reported that functionalization, doping, and defects can tune the electronic and magnetic properties of the nanomaterials nanomaterials [13-24]. Motivated by the successful detection of individual gas molecules by graphene, the sensing capability of other two-dimensional (2D) structures toward different polluting gasses have been explored [25]. Unlike flat graphene sheet, silicene and germanene have buckled honeycomb structures due to the partial sp[3] hybridization of Si and Ge atoms [26, 27], making them chemically more reactive toward atoms and molecule adsorption compared to graphene [28-31]. Although N-based molecules are chemisorbed on silicene and germanene *via* strong covalent bonds, $CO_2$ is weakly physisorbed on silicene and germanene sheets [29, 30, 32-35]. The electronic structures of silicene and germanene show strong modifications under Li decoration [36-39]. It was also stated that functionalization of Si atoms with Au (Li) improves the interaction between silicene and CO ($CO_2$) molecules [40, 41]. Yuan *et al.* performed DFT calculations to investigate the stability, structural and electronic properties of saturated and half-saturated germanene with alkali metal atoms and found that the complete lithiated germanene has the highest stability among all the studied structures [42].

In this study, density functional theory (DFT) method is accepted to examine the $CO_2$ adsorption on germanene sheet functionalized by Li atoms.

**2 Computational Method** Calculations are performed using first-principle methods based on DFT implemented in Atomistix ToolKit (ATK) package [43]. The exchange-correlation functional is approximated by the Generalized Gradient Approximation of Perdew-Burke-Ernzerhof (GGA-PBE) with a double-ζ polarized basis set. To describe long-range van der Waals (vdW) interactions, the Grimme vdW correction (DFT-D2) [44] is also considered. The density mesh cut-off for plane-wave expansion is set to be 150 Ry. To avoid adjacent images interactions, a large vacuum space of 25 Å is considered in *z*-direction. Prior to the calculations; all the structures are fully relaxed using the conjugate gradient method up until the force on each atom is less than 0.01 eV/Å. For the germanene unit cell, the first Brillouin zones are sampled using 11×11×1 and 21×21×1 *k*-points for optimization and calculations, respectively.

**3 Results and discussions** We first discuss the adsorption behavior of $CO_2$ gas molecules on the pristine germanene sheet. To this end, an individual $CO_2$ is initially placed on a germanene sheet at four different positions including valley (the Ge atom in the lower sublattice), hill (the Ge atom in the upper sublattice), bridge (the Ge-Ge bond), and hollow (the center of a hexagon ring) sites with two different molecular orientations, parallel and perpendicular to the surface. We use the notation $Ge_8(CO_2)_n$ to distinguish between the various numbers of $CO_2$ molecules, *n*, in a 2×2 supercell of germanene (adopted from Ref. [41]). The adsorption behavior of molecule on germanene is investigated after full relaxation. The structural stability can be addressed using adsorption energy $(E_{ad})$ which is

$$E_{ad} = [E_{Germanene+CO_2} - E_{Germanene} - n \times E_{CO_2}]/n \quad (1)$$

Here, $E_{Germanene+CO2}$, $E_{Germanene}$, and $E_{CO2}$ denote the total energies of the germanene-$CO_2$ system, pristine germanene, and the isolated $CO_2$ molecule, respectively. Based on the definition, the negative $E_{ad}$ represents the structural stability. Moreover, for a material to be suitable as a medium for $CO_2$ capture, relatively large amounts of adsorption energy and charge transfer are vital. Comparing the adsorption energies of different adsorption geometries, the horizontal alignment on top of the bridge site is found to be energetically more favorable, as shown in Fig. 1(a). The distance between $CO_2$ and the germanene sheet is 3.88 Å, showing weak vdW interactions. A small adsorption energy of −0.11 eV along with small charge transfer (calculated by Mulliken population analysis) of 0.04 *e* from the molecule to the germanene confirms this fact that $CO_2$ molecule is physisorbed on the germanene. Therefore, pristine germanene could not be an appropriate material for $CO_2$ capture because minimum adsorption energy of −0.14 eV is required from an application point of view [45]. These results agree well with previous theoretical [29]. It should be noted that for higher $CO_2$ coverage, the molecules are inclined with respect to the horizontal direction due to the repulsion between them, as shown in Fig. 1(c). The distances between $CO_2$ molecules and germanene are 3.51 to 3.97 Å. The average adsorption energy and charge transfer from each $CO_2$ to germanene are decreased to −0.09 eV and 0.02 *e* for high coverage, respectively. These small values limit the application of germanene as a potential media to capture $CO_2$. Pristine germanene is a zero-gap semiconductor. We found that the linear Dirac-like dispersion relation of germanene at K point remains almost unchanged upon physisorption of a single $CO_2$ gas molecule. A tiny band gap of 3 meV is opened at the Dirac point of germanene, as shown in Fig. 1(b). The change in the band structure of germanene is a little more pronounced at the high coverage of $CO_2$ where the band gap is enhanced to 44 meV, as illustrated in Fig. 1(d). I

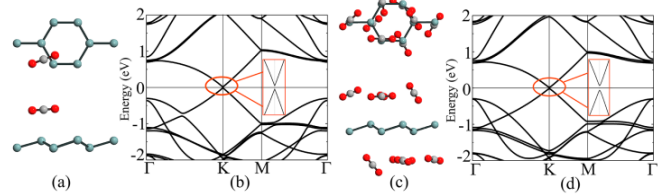

**Figure 1** The most stable adsorption configurations (top and side view) and their corresponding band structures for (a) and (b) one $CO_2$ (low coverage) (c) and (d) eight $CO_2$ (high coverage) on pristine germanene, respectively. The cyan, gray, and red balls represent Ge, C, and O atoms, respectively.

Lithium functionalization can enhance the stability of $CO_2$ gas molecules on germanene nanosheet [42]. Similar to fully lithiated silicene [46] and graphene [47], lithium atoms occupy sites on top of the Ge atoms in the lower sublattice, which is in agreement with previous findings [42]. The optimized geometry of Li-functionalized germanene is presented in Fig. 2(a). The lattice distortion caused by lithium adsorption is noticeable. The buckling distance and Ge-Ge bond length are enlarged from 0.73 and 2.46 Å in pristine germanene to 1.23 and 2.55 Å in lithiated germanene, respectively. Furthermore, the minimum Ge-Li bond length is 2.55 Å. The bond strengths can be assessed on the basis of the adsorption energy of lithium atoms on the surface of germanene using following formula:

$$E_{ad} = [E_{Germanene+Li} - E_{Germanene} - N_{Li} \times E_{Li}]/N_{Li} \quad (2)$$

Here, $E_{Germanene+Li}$ and $E_{Li}$ are the total energies of the lithiated germanene system and the single Li atom, respectively. In addition, $N_{Li}$ is the number of adsorbed Li atoms in the supercell. Adsorption energy of −1.01 eV/atom reflects the stability of fully lithiated germanene ($Ge_8Li_8$) and suggests the chemisorption of lithium atoms on the surface of germanene. However, the electronic total charge density indicates that there is a small electron orbital overlap between Li and Ge atoms, showing that a weak covalent bonding exists between Li and germanene, see Fig. 2(c). It is found that each Li atom (electronegativity of 1) donates 0.176 *e* to the more electronegative Ge atoms (electronegativity of 2); hence, the main character of the bonding between Ge-Li in this system is ionic. Consequently, the ionic interaction induced by large charge transfer between Li and Ge atoms is the reason of high stability of fully lithiated germanene. Upon complete lithiation, the band structure is transformed from a zero-gap semiconductor to an indirect semiconductor with 220 meV band gap, as shown in Fig. 2(b). Unlike pristine germanene, fully lithiated germanene does not have the linear dispersion at the K point.

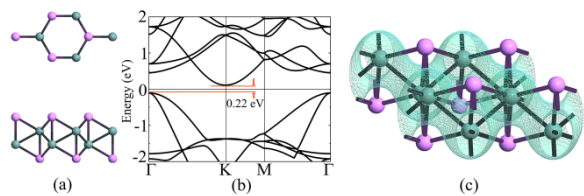

**Figure 2** (a) Relaxed configuration (top and side view) of fully lithiated germanene, its corresponding (b) band structure, and (c) electronic total charge density distributions.

Osborn *et al.* predicted an energy gap of 368 meV in fully lithiated silicene [46]. A Ge atom has a larger atomic radius in comparison with Si atom, giving rise to a decrease in the strength of covalent bonds in Ge-Li compared to Si-Li. Hence, larger energy band gap in fully lithiated silicene than that in germanene can be associated with the stronger covalent nature in the former. Moreover, it has been reported that graphene behaves as a metal upon complete lithiation and keeps its linear dispersion at the K point [47].

Next, we study the adsorption of $CO_2$ molecules on Li-functionalized germanene ($Ge_8Li_8(CO_2)_n$). For a single $CO_2$ molecule adsorption ($Ge_8Li_8(CO_2)_1$), the most energetically favorable position was found to be on the top of Ge atom in the upper sublattice (hill site) with an adsorption energy of −2.31 eV, showing strong interactions between the molecule and the Li-functionalized germanene, as shown in Fig. 3(a). The triatomic molecule of $CO_2$ with 180° bond angles is strongly distorted with the O-C-O bond angle of 123.3°. The C atom points to the germanene sheet and C-Ge bond length (the distance between the molecule and Li-functionalized germanene sheet) is 2.13 Å which is much smaller than that of pristine germanene sheet (3.88 Å). The $CO_2$ molecule also interacts with Li atoms by its O atoms, where Li-O bonds are 1.85 and 1.75 Å. Also, the C-O bonds (1.18 Å) in an isolated $CO_2$ molecule get elongated to 1.28-1.30 Å in $Ge_8Li_8(CO_2)_1$. The charge redistribution between germanene, Li, and $CO_2$ (concerning the isolated Li-functionalized germanene and $CO_2$) are calculated and listed in Table 1. Li-functionalized germanene donates 0.338 *e* electrons to $CO_2$. In comparison with low electrons transfer (−0.04 e) in the case of non-functionalized germanene, $CO_2$ is significantly stabilized on Li-functionalized germanene due to the large charge transfer.

With increasing the $CO_2$ molecules in the supercell, the adsorption energy decreases toward a minimum of −2.60 eV (50% $CO_2$ coverage) and increases again (less stability), see Fig. 3(c). For high $CO_2$ coverage (100% coverage), the molecules are positioned on top of Li atoms in an average binding distance of 2 Å which is much smaller than non-functionalized germanene (3.51-3.97 Å). Only a part of $CO_2$ molecules (O atom) is firmly bound to the Li-functionalized germanene and an inclination in the range of 36-48° with respect to the vertical direction of the germanene sheet is found for the $CO_2$ molecules due to the repulsion between the molecules, as shown in Fig. 3(d). The moderate adsorption energy of −0.80 eV and a small charge transfer of 0.047 *e* suggests a weak interaction that makes the release step of $CO_2$ molecules easier, while stabilizes $CO_2$ on the adsorbent. If another $CO_2$ molecule is added into the supercell, the adsorption is even further increased to −0.22 eV. The $CO_2$ capacity is calculated to be 12.57 mol/kg. It should be noted that the $Ge_8Li_8(CO_2)_n$ structure can be recovered to its original form after removal of all $CO_2$ molecules.

**Table 1** The adsorption energy $E_{ad}$ (eV) per molecule, energy band gap $E_g$ (eV), and the charge redistribution on germanene, Li atoms, and each $CO_2$.

|  | $E_{ad}$ (eV) | $E_g$ (eV) | $Q_{MPA}$ (*e*) on Germanene | Li | $CO_2$ |
|---|---|---|---|---|---|
| $Ge_8(CO_2)_1$ | −0.11 | 0.003 | 0.041 | - | −0.04 |
| $Ge_8(CO_2)_8$ | −0.09 | 0.044 | 0.176 | - | −0.02 |
| $Ge_8Li_8$ | −1.01 | 0.220 | 1.408 | −1.408 | - |
| $Ge_8Li_8(CO_2)_1$ | −2.31 | 0.520 | −0.564 | 0.2660 | 0.338 |
| $Ge_8Li_8(CO_2)_2$ | −2.48 | 0.620 | −0.343 | −0.367 | 0.354 |
| $Ge_8Li_8(CO_2)_3$ | −2.51 | 0.790 | −0.113 | −0.848 | 0.321 |
| $Ge_8Li_8(CO_2)_4$ | −2.60 | 1.070 | 0.421 | −1.876 | 0.364 |
| $Ge_8Li_8(CO_2)_5$ | −2.51 | 0.960 | 0.337 | −2.090 | 0.350 |
| $Ge_8Li_8(CO_2)_6$ | −1.70 | 0.640 | 0.645 | −2.701 | 0.342 |
| $Ge_8Li_8(CO_2)_7$ | −1.54 | 0.228 | 0.438 | −1.623 | 0.378 |
| $Ge_8Li_8(CO_2)_8$ | −0.80 | 0.000 | 0.700 | −1.709 | 0.047 |
| $Ge_8Li_8(CO_2)_9$ | −0.22 | 0.000 | −0.213 | −0.015 | −0.02 |

The band structures of Li-functionalized germanene after $CO_2$ adsorption with low and high coverage are also shown in Fig. 3(b) and (e), respectively. The band gap of the Li-functionalized germanene increases from 0.520 eV (12.5% $CO_2$ coverage) toward a maximum of 1.070 eV (50% $CO_2$ coverage), and decreases to zero at 100% coverage, see Fig. 3 (f). Therefore, the $CO_2$ coverage can be evaluated using the band gap because of the high sensitivity of Li-functionalized germanene toward $CO_2$ adsorption.

**4 Conclusions** Calculations based on DFT method have been performed to explore the capability germanene toward $CO_2$ adsorption. Our results revealed that the interaction of a $CO_2$ gas molecule with pristine germanene is very weak due to the small charge transfer. However, the adsorption energy of the Li-functionalized germanene is significantly higher than that of pristine for low $CO_2$ coverage. At high $CO_2$ coverage, the interactions of $CO_2$ molecules with Li-functionalized germanene decrease due to the repulsion between molecules. Nevertheless, the adsorption energy is enough to stabilize the $CO_2$ on the adsorbent. Moreover, the structure of Li-functionalized germanene can be recovered to its initial shape after removal of $CO_2$ molecules. A large $CO_2$ storage capacity of 12.57 mol/kg is predicted for the system, and the band gap of the system is highly sensitive to the $CO_2$ coverage.

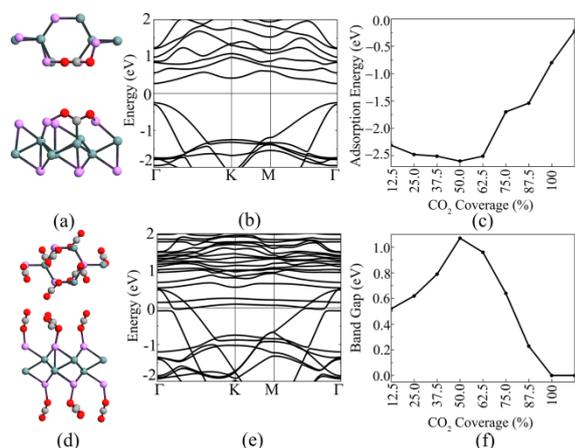

**Figure 3** The most stable adsorption configurations (top and side view) and their corresponding band structures of (a) and (b) $Ge_8Li_8(CO_2)_1$ (low coverage) (d) and (e) $Ge_8Li_8(CO_2)_8$ (high coverage). The adsorption energy and band gap changes of $Ge_8Li_8(CO_2)_n$ as a function of $CO_2$ coverage.